\documentclass{PoS}
\usepackage{epsfig}
\usepackage{multicol}
\usepackage{graphicx}
\usepackage{here}
\PoS{PoS(HEP2005)}

\newcommand{\bea}{\begin{eqnarray}}
\newcommand{\eea}{\end{eqnarray}}

\def \beq{\begin{equation}}
\def \eeq{\end{equation}}

\def \bit{\begin{itemize}}
\def \eit{\end{itemize}}

\def \a{\alpha}

\def \m{\mu}

\title{Towards the NNLL precision in the decay $\bar B \rightarrow X_s \gamma$}

\ShortTitle{Towards the NNLL precision in $\bar B \rightarrow X_s \gamma$}

\author{Hrachia M.  Asatrian, Artyom Hovhannisyan, and Vahagn Poghosyan\\
Yerevan Physics Institute, 375036, Yerevan, Armenia}

\author{Christoph Greub\\
Institute for Theoretical Physics, University Berne, CH-3012 Berne, Switzerland}

\author{\speaker{Tobias Hurth}\thanks{Heisenberg Fellow}\\
        CERN, Dept.\ of Physics, Theory Unit, CH-1211 Geneva 23, Switzerland\\ SLAC, Stanford University, Stanford, CA 94309, USA}

\abstract{The present NLL prediction for the decay rate of the rare inclusive
process $\bar B \rightarrow X_s \gamma$ has a large uncertainty due to 
the charm mass renormalization scheme ambiguity.
We estimate that this uncertainty will be reduced by a factor of 2  at the 
NNLL level. 
This is a strong motivation for the on-going NNLL 
calculation, which will thus significantly  increase the sensitivity of the 
observable $\bar B \rightarrow X_s \gamma$ to possible 
new degrees  of freedom beyond the SM. 
We also give a brief status report of the NNLL calculation.}

\FullConference{HEP2005 International Europhysics 
                Conference on High Energy Physics\\
                21--27 July 2005\\
	        Lisbon, Portugal\\
CERN-PH-TH/2005-233, SLAC-PUB-11586}

\begin{document}

The inclusive decay $\bar B \rightarrow X_s \gamma$ is well known as 
one of the  most important flavour observables within the indirect search 
for new physics \cite{Hurth:2003vb}. 
The  present  experimental accuracy already reached the $10 \%$ level, as
reflected in the world average of the present
measurements \cite{HFAG}:
\beq
\mbox{BR}[\bar B \to X_s \gamma] = (3.39^{+0.30}_{-0.27}) \times 10^{-4} \,.
\label{world}
\eeq
In the near future, more precise data on this mode are expected from
the $B$-factories. Thus, it is mandatory to reduce the present theoretical 
uncertainty accordingly. Non-perturbative effects are naturally small 
within inclusive modes \cite{Hurth:2003vb}; 
also additional non-perturbative corrections due to necessary cuts in the 
photon energy spectrum are under control (see \cite{Neubert}).
As was first noticed in \cite{Gambino}, there exists a large 
uncertainty in the theoretical NLL prediction related to the renormalization 
scheme of the charm-quark mass on which we focus in this article. 
The reason is that the matrix elements 
$\langle s \gamma |{O}_{1,2}| b \rangle$, 
through which the charm-quark mass dependence dominantly enters,
vanish at the lowest order (LL) and, as a consequence, 
the charm-quark mass does not get renormalized in a  NLL calculation,
which means that the symbol $m_c$ can be identified with $m_{c,\rm{pole}}$ or
with the $\overline{\mbox{MS}}$ mass $\bar{m}_c(\mu_c)$ at some scale $\mu_c$
or with some other definition of $m_c$.
In a recent theoretical update of the NLL prediction of this branching
fraction the ratio
$m_c/m_b$ was varied in the conservative range $0.18 \le m_c/m_b \le 0.31$
that  covers both the pole mass value (with its numerical error) and the 
running mass $\bar{m}_c(\mu_c)$ value (with $\mu_c \in [m_c,m_b]$), leading to 
\cite{Hurth:2003dk}:
\begin{equation}
\label{hurth_lunghi}
\mbox{BR}[\bar{B} \to X_s \gamma] = (3.70 \pm 0.35|_{m_c/m_b} \pm
0.02|_{\rm{CKM}} \pm 0.25|_{\rm{param.}} \pm 0.15|_{\rm{scale}}) \times 10^{-4} \, .
\end{equation}
The only way to resolve this scheme ambiguity in a satisfactory way
is to perform a systematic NNLL calculation.
Working to next-to-next-to-leading-log 
(NNLL) precision means that one is resumming all the terms of the form 
\begin{equation}      
(\alpha_s(m_b))^p  \, \alpha_s^n(m_b) \, \log^n (m_b/M) \, , \quad (p=0,1,2).
\label{NLLQCD}
\end{equation}
where $M=m_t$ or $M=m_W$, $n=0,1,2,...\,\,  $ .
Such a calculation is most suitably done in the framework of
an  effective 
low-energy theory.
The effective interaction Hamiltonian can be written as
\begin{equation}
 H_{\rm{eff}} = - {4 G_{F}}/{\sqrt{2}} \,\,\,\,\,\,\, V_{tb} V_{ts}^* \, 
\sum   {C_{i}(\mu, M)}\,\,    {O}_i(\mu) \, , 
\end{equation}
where ${O}_i(\m)$ are the relevant dimension 6 operators
and 
$C_{i}(\mu, M)$ are the Wilson coefficients.

Parts of the  three principal calculational 
steps leading to the NNLL result within the
effective field theory approach  are already done:
{\bf (a)} The full SM theory has to be matched 
with the effective theory at the scale $\m=\mu_W$, where
$\mu_W$ denotes a scale of order $m_W$ or $m_t$. 
The Wilson coefficients 
$C_i(\mu_W)$ only pick up small QCD corrections,
which can be calculated in fixed-order perturbation theory. 
In the NNLL  program, the matching has to be worked out at the 
order $\a_s^2$. The matching calculation to this precision is already
finished, including the most difficult piece, the three-loop matching 
of the operators ${O}_{7,8}$ \cite{Misiak}.
{\bf (b)} The  evolution of these Wilson 
coefficients from  $\m=\mu_W$ down to $\m = \mu_b$ then has to be performed 
with the help of the renormalization group, where $\mu_b$ is of the 
order of $m_b$.
As the matrix elements of the operators evaluated at the low scale
$\mu_b$ are free of large logarithms, the latter are contained in resummed
form in the Wilson coefficients. For the NNLL  calculation, this RGE step
has to be done  using the anomalous--dimension matrix up  to order $\a_s^3$. 
While the three-loop mixing among the four-quark operators 
${O}_i$ $(i=1,\ldots, 6)$ \cite{GorbahnHaisch}
and among the dipole operators ${O}_{7,8}$ \cite{GorbahnHaischMisiak}
are already available, the four-loop mixing of the four-quark into the dipole
operators is still an open issue. {\bf (c)} 
To achieve NNLL  precision, the matrix elements 
$\langle X_s \gamma  |{O}_i (\mu_b)|b \rangle$ have to be calculated 
to order $\a_s^2$ precision. This includes also bremsstrahlung corrections. 
In 2003, the ($\alpha_s^2\,n_f$) corrections to the matrix elements of the
operators $O_1$,$O_2$,$O_7$,$O_8$ were calculated \cite{Bieri}. 
Complete order $\alpha_s^2$ results are available to the $(O_7,O_7)$
contribution to the decay width \cite{Blokland:2005uk}.
Recently, also order $\alpha_s^2$ terms
to the photon energy spectrum (away from the endpoint $E_\gamma^{\rm{max}}$)
were worked out for the operator $O_7$ \cite{Melnikov:2005bx}.

\begin{figure}[tbph]
\begin{center}
\epsfig{file=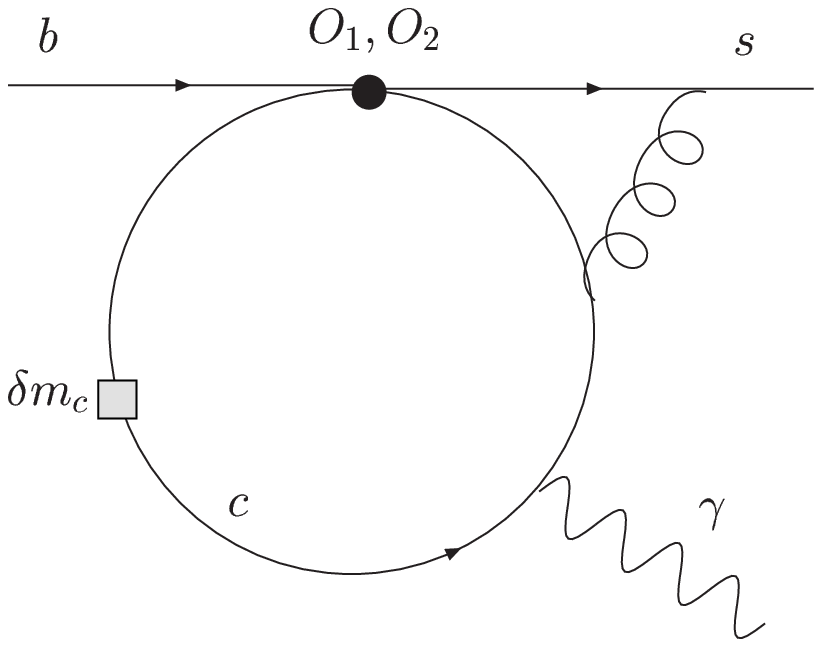, width=4.0cm}\epsfig{file=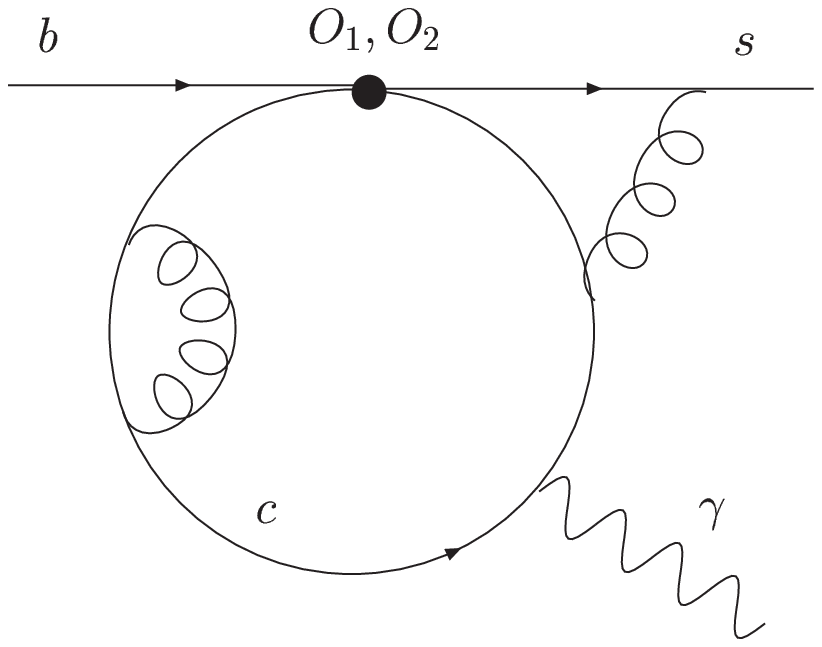, width=4.0cm}
\end{center}
\caption[]{Left: 
    Typical $\delta m_c$ insertion diagram. Right: 
    Typical diagram with a self energy  insertion.}
\label{fig:diag}
\end{figure}

In ref. \cite{Asatrian} a strong motivation for this complicated NNLL
effort was given by calculating those NNLL terms that  are induced by
renormalizing the charm-quark mass in the NLL expressions, i.e. those
terms that  are sensitive to the definition of the charm-quark mass.
These terms correspond to $\delta m_c$ insertions in the diagrams
related to the NLL order matrix elements
$M_{1,2}^{\rm{virt}}(m_c) = \langle s \gamma|{O}_{1,2}(\mu_b)|b \rangle$ 
and $M_{1,2}^{\rm{brems}}(m_c)=\langle s \gamma g|{O}_{1,2}(\mu_b)|b \rangle$
(for an example, see the left diagram in  Fig. \ref{fig:diag}).

The sum 
$\delta M^{\rm{virt} (\epsilon)}_{1,2}(m_c) 
\cdot \delta m_c$ of all these
insertions can be obtained by replacing $m_c$ by $m_c + \delta m_c$ in the
$O(\alpha_s^1)$ results $M^{\rm{virt}(\epsilon)}_{1,2}(m_c)$, followed by expanding in $\delta
m_c$ up to linear order:
\begin{equation}
\label{massshift}
M_{1,2}^{\rm{virt}(\epsilon)}(m_c+\delta m_c) = M_{1,2}^{\rm{virt}(\epsilon)}(m_c)+
\delta M^{\rm{virt}(\epsilon)}_{1,2}(m_c) \cdot \delta m_c +
O((\delta m_c)^2) \, .
\end{equation} 
As $\delta m_c$ is ultraviolet-divergent, the matrix elements $M_{1,2}^{\rm{virt}(\epsilon)}(m_c)$ 
are needed in our application up to order $\epsilon^1$, 
as indicated by the notation in eq. (\ref{massshift}). 
In \cite{Asatrian} the explicit analytical results for these matrix elements 
are given in such a way that they can be used in a future
complete  NNLL calculation.  
The explicit shift $\delta m_c$ depends of course on the renormalization
scheme. When aiming at expressing the results for 
$M_{1,2}^{\rm{virt}(\epsilon)}(m_c)$ in terms
of $\bar{m}_c(\mu_b)$ or  $m_{c,\rm{pole}}$, the shift reads ($C_F=4/3$)
\begin{eqnarray}
\label{mcshift_bar}
\delta \bar{m}_{c}(\mu_b) = -\frac{\alpha_s(\mu_b)}{4\pi} \, C_F \,
\frac{3}{\epsilon} \, \bar{m}_{c}(\mu_b)\,\,\,\,  \mbox{or} \,\,\,\,   
\delta m_{c,\rm{pole}} = -\frac{\alpha_s(\mu_b)}{4\pi} \, C_F \, \left( \frac{3}{\epsilon}
  + 3 \ln \frac{\mu_b^2}{m_c^2} + 4 \right) \, m_{c,\rm{pole}}\,. 
\nonumber
\end{eqnarray}
The infinities induced by the $1/\epsilon$ terms in $\delta m_c$ get cancelled
in a full NNLL calculation, in particular by self-energy diagrams depicted in the right diagram in 
  Fig. \ref{fig:diag}. When implementing these self-energy insertions,
we only took into account the  
$\Sigma_1(p^2=m_c^2)$ piece, i.e.
that part of the one-loop self-energy which only gets renormalized 
by the mass parameter. When used at the fixed momentum $p^2=m_c^2$, this 
piece is gauge-independent. 

\begin{figure}[tbph]
\begin{center}
\epsfig{file=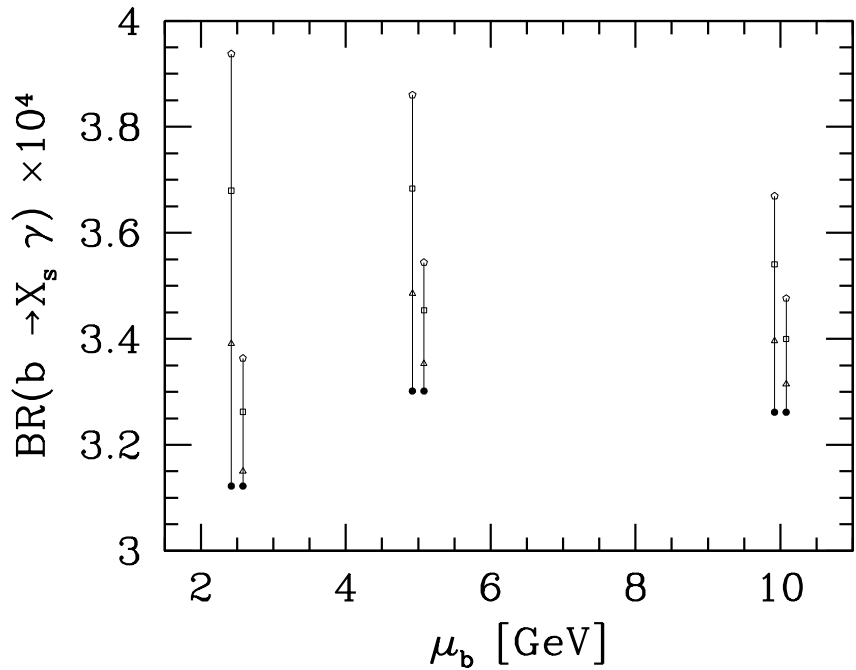,width=12.5cm}
 \caption{$\mbox{BR}(b \to X_s \gamma)$ for three values of $\mu_b$ (see text for more details).}
    \label{fig:plots}
    \end{center}
\end{figure}

Our final estimates are given in Fig. \ref{fig:plots} 
for three different values of $\mu_b$, where $\mu_b$ represents the 
usual renormalization scale of the effective field theory.  
Within each vertical string, the solid dot 
represents the branching ratio using 
the pole mass  $m_{c,\rm{pole}}$, while the open symbols correspond 
to the $\overline{\mbox{MS}}$ mass $\bar{m}_c(\mu_c)$ 
for $\mu_c=1.25$ GeV (triangle), 
    $\mu_c=2.5$ GeV (quadrangle) and
    $\mu_c=5.0$ GeV (pentagon). 
For each $\mu_b$ the left string shows the value of
the branching ratio at the NLL level, while the right string shows  
the corresponding value where, in addition. $\delta m_c$ mass insertions and 
$\Sigma_1(p^2=m_c^2)$ insertions were taken into account. 
Because the combination of these insertions is zero 
by construction for the pole scheme, 
the solid dots are at the same place in the left and the right string 
for a given value of $\mu_b$. 
We stress that all the statements made in the following 
are independent of this absolute normalization introduced 
by the  additional $\Sigma_1(p^2=m_c^2)$ insertions, because
we refer to the {\it reduction of the error} only.
From Fig. \ref{fig:plots} we see that the error related to the charm-quark 
mass definition is significantly reduced when 
the NNLL terms connected with mass insertions are taken into account. Taking as an 
example the results for $\mu_b=5$ GeV, 
we find that at the NLL level the
branching ratio evaluated for $\bar{m}_c(2.5 \ \mbox{GeV})$
is $12.6\%$ higher than the one based on
$m_{c,\rm{pole}}$.  Including the new contributions, these $12.6\%$
get reduced to $5.1\%$. One also can read off an analogous 
significant reduction within the $\overline{\mbox{MS}}$  scheme itself. 
However, to obtain a NNLL prediction for the central value
of the branching ratio, it is of course necessary to calculate all NNLL terms. 
\vspace{-3cm}
\newpage

\frenchspacing
\footnotesize
\begin{multicols}{2}

\end{multicols}

\begin{thebibliography}{99}
\footnotesize{
\bibitem{Hurth:2003vb}
T.~Hurth,
Rev. Mod. Phys. {\bf 75} (2003) 1159
[arXiv:hep-ph/0212304].
\bibitem{HFAG}
  Heavy Flavor Averaging Group (HFAG),\\
http://www.slac.stanford.edu/xorg/hfag/
\bibitem{Neubert}
  M.~Neubert,
  Eur.\ Phys.\ J.\ C {\bf 40} (2005) 165
  [arXiv:hep-ph/0408179].
\bibitem{Gambino}
P.~Gambino and M.~Misiak,
Nucl.\ Phys.\ B {\bf 611} (2001) 338
[arXiv:hep-ph/0104034].
\bibitem{Hurth:2003dk}
T.~Hurth, E.~Lunghi and W.~Porod,
Nucl.\ Phys.\ B {\bf 704} (2005) 56
[arXiv:hep-ph/0312260].
\bibitem{Misiak}
  M.~Misiak and M.~Steinhauser,
  Nucl.\ Phys.\ B {\bf 683} (2004) 277
  [arXiv:hep-ph/0401041].
\bibitem{GorbahnHaisch}
 M.~Gorbahn and U.~Haisch,
 Nucl.\ Phys.\ B {\bf 713} (2005) 291
 [arXiv:hep-ph/0411071].
\bibitem{GorbahnHaischMisiak}
  M.~Gorbahn, U.~Haisch and M.~Misiak,
  Phys.\ Rev.\ Lett.\  {\bf 95} (2005) 102004
  [arXiv:hep-ph/0504194].
\bibitem{Bieri}
  K.~Bieri, C.~Greub and M.~Steinhauser,
  Phys.\ Rev.\ D {\bf 67} (2003) 114019
  [arXiv:hep-ph/0302051].
\bibitem{Blokland:2005uk}
  I.~Blokland, A.~Czarnecki, M.~Misiak, M.~Slusarczyk and F.~Tkachov,
  Phys.\ Rev.\ D {\bf 72} (2005) 033014
  [arXiv:hep-ph/0506055].
\bibitem{Melnikov:2005bx}
  K.~Melnikov and A.~Mitov,
  Phys.\ Lett.\ B {\bf 620} (2005) 69
  [arXiv:hep-ph/0505097].
\bibitem{Asatrian}
  H.~M.~Asatrian, C.~Greub, A.~Hovhannisyan, T.~Hurth and V.~Poghosyan,
  Phys.\ Lett.\ B {\bf 619} (2005) 322
  [arXiv:hep-ph/0505068].
}
\end{thebibliography}
\end{document}